\documentclass[10pt,letterpaper]{article}
\usepackage{opex3}
\usepackage{ulem}

\begin{document}

\title{250-MHz synchronously pumped optical parametric
oscillator at \\ 2.25-2.6~$\mu$m and 4.1-4.9~$\mu$m}

\author{Nicola Coluccelli,$^{1,*}$ Helge Fonnum,$^2$  Magnus Haakestad,$^2$ \\ Alessio Gambetta,$^1$
Davide Gatti,$^1$ Marco Marangoni,$^1$ \\ Paolo Laporta,$^1$ and
Gianluca Galzerano$^1$}

\address{
$^1$Dipartimento di Fisica - Politecnico di Milano and Istituto di
Fotonica e Nanotecnologie - CNR, Piazza Leonardo da Vinci 32,
20133 Milano, Italy,\\
$^2$FFI (Norwegian Defence Research Establishment), PO Box 25,
NO-2027 Kjeller, Norway }

\email{$^*$nicola.coluccelli@polimi.it} 



\begin{abstract}A compact and versatile femtosecond mid-IR source is presented, based on an
optical parametric oscillator (OPO) synchronously pumped by a
commercial 250-MHz Er:fiber laser. The mid-IR spectrum can be tuned
in the range 2.25-2.6~$\mu$m (signal) and 4.1-4.9~$\mu$m (idler),
with average power from 20 to 60~mW. At 2.5~$\mu$m a minimum pulse
duration of 110~fs and a power of 40~mW have been obtained. Active
stabilization of the OPO cavity length has been achieved in the
whole tuning range.
\end{abstract}




\section{Introduction}
\noindent The near-mid infrared spectral region from 2 to 10~$\mu$m
is of primary importance in molecular spectroscopy, due to the
strong absorption provided by the fundamental roto-vibrational bands
and first overtones of many molecules. In this context, the
availability of stable and accurate probes in the mid-IR spectral
window is greatly desirable for a wide range of applications,
including air quality and environment monitoring, industrial
processes control, human breath analysis and detection of
biologically hazardous or explosive materials. State of the art
performances have been reached by exploiting phase-stabilized mid-IR
femtosecond comb sources in direct multiplex spectroscopy
experiments, where high-sensitivity and parallel detection of
multiple molecular species have been simultaneously
demonstrated~\cite{Diddams}. Different approaches have been used to
extend frequency combs in the mid-infrared. Recently, a femtosecond
Tm:fiber source has been proposed to cover the range 1-2.5~$\mu$m
with a power per mode below 10~nW exploiting nonlinear broadening in
a silica fiber~\cite{Jiang}. Mode-locked (ML) Cr:ZnSe lasers are
good candidates in the region from 2.2 to 2.8~$\mu$m, but their
emission is presently limited to a 70-nm band around
2.4~$\mu$m~\cite{Cizmeciyan}. Various systems based either on
difference frequency generation~\cite{Neely,Ruehl} or on
OPOs~\cite{Adler,Leindecker,Washburn,Leindecker2,Ferreiro1} have
been also reported at wavelengths above 3~$\mu$m, with power levels
as high as 1.5~W~\cite{Adler} and ultrawide spectral coverage over
the range 2.6-6.1~$\mu$m~\cite{Leindecker}. However, no mid-IR
fiber-pumped synthesizer has been demonstrated so far at repetition
frequencies higher than 250~MHz, which would be very useful for
direct comb spectroscopy techniques and, in particular, for
dual-comb spectroscopy \cite{Keilmann,Bernhardt,Coddington}.

In this paper, we describe a robust OPO synchronously pumped by a
commercial Er:fiber femtosecond laser, operating at a repetition
frequency of 250~MHz. The OPO signal and idler have been tuned in
the range 2.25-2.6~$\mu$m and 4.1-4.9~$\mu$m, respectively, with
average powers from 20 to 70~mW; this corresponds to a power per
mode above 1~$\mu$W at all wavelengths, i.e. above the minimum power
level generally required in gas trace detection experiments, which
makes this system especially suitable to cover the lack of frequency
comb sources for molecular spectroscopy in the region from 2.2 to
2.6~$\mu$m. In a previous work based on a similar OPO system, power
levels of $\sim$30~mW in the 3.7-4.7~$\mu$m spectral range were
obtained at a repetition frequency of 90~MHz~\cite{Haakestad}. Here,
we report on a completely different design of the OPO cavity,
exploiting a ring configuration to achieve unidirectional operation
and thus to minimize the losses. Particular attention has been
devoted to the optimization of the OPO signal at 2.25-2.6~$\mu$m,
where we were able to obtain pulses with a minimum duration of
110~fs at 2.5~$\mu$m and an average power of $\sim$40~mW. In
addition, long term intensity stabilization of the signal in the
whole tunability range from 2.25 to 2.6~$\mu$m has been achieved by
locking the OPO resonator length, which makes this system suitable
for direct comb spectroscopy and ultrafast spectroscopy applications
in the mid-IR. Once aligned, our OPO system is very robust and
guarantees hands-free operation for hours.

\section{OPO experimental setup}

\begin{figure}[t]
\centerline{\includegraphics[width=7.5cm]{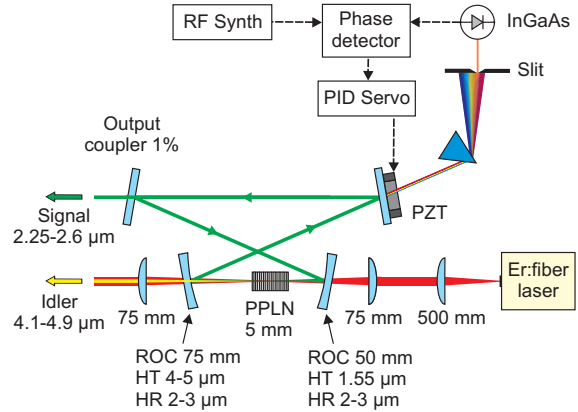}}
\caption{\label{setup}Scheme of the OPO cavity and setup for offset
frequency locking.}
\end{figure}

\begin{figure}[b]
\centerline{\includegraphics[width=10cm]{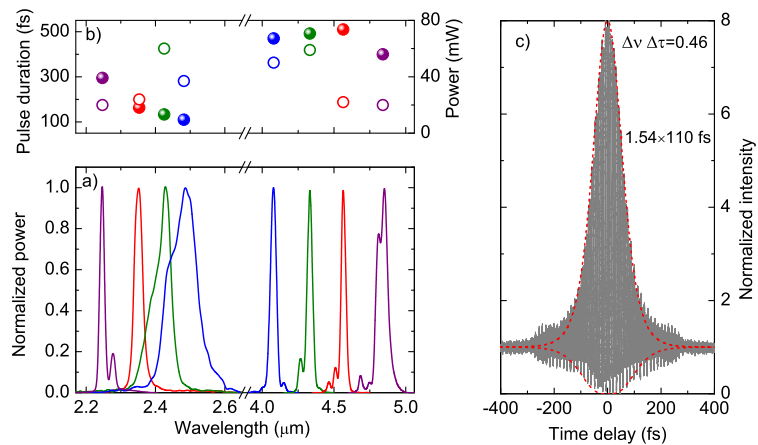}}
\caption{\label{spectra} (a) Spectra of the OPO signal and idler as
acquired with poling period of 29.5~$\mu$m (violet), 30.5~$\mu$m
(red), 31.5~$\mu$m (green), 32.5~$\mu$m (blue). (b) Pulse duration
(filled circles) and power level (open circles) of the OPO signal
and idler.  (c) Autocorrelation of the 110-fs OPO signal pulses for
a poling period of 32.5~$\mu$m.}
\end{figure}

\noindent The scheme of the OPO cavity, pump system and locking
electronics is shown in Fig.~\ref{setup}. The pump source is a
250-MHz ML Er:fiber laser (MenloSystems, FC1500) with output power
of 580~mW, pulse duration of 40~fs, and vertical linear polarization
on the fundamental mode. The pump beam is first collimated by an
anti-reflection (AR) coated lens with focal length of 500~mm and
then focused onto a periodically poled 5\% doped MgO:LiNbO$_3$
(PPLN) crystal by an AR coated lens with focal length of 75~mm. The
calculated pump beam waist inside the PPLN crystal is 24~$\mu$m. The
OPO cavity is constituted by four mirrors in a ring configuration
with an unfolded length of $\sim$1.2~m: a curved mirror with radius
of curvature (ROC) of 50~mm and a dichroic coating (AR at
1.55~$\mu$m, high-reflection(HR)$>$99.8\% at 2-3~$\mu$m), a curved
mirror with ROC of 75~mm and a dichroic coating for extraction of
the idler (AR at 4-5~$\mu$m, HR at 2-3~$\mu$m, YAG substrate), a
plane HR mirror and an output coupler (R$=$99\%) for the 2-3~$\mu$m
range. The calculated mode radius at the intracavity focus (signal
wavelength) in the PPLN crystal is $\sim$27~$\mu$m, with good
matching to the pump beam. The 5-mm long PPLN crystal, located at
the intracavity focus of the signal mode, has six different poling
period from 29.5 to 34.5~$\mu$m with steps of 1~$\mu$m, and is AR
coated with R$<$4\% at the pump, signal, and idler wavelengths. The
PPLN crystal is mounted inside an oven to allow for a fine tuning of
the OPO emission frequency; for brevity temperature tuning curves
are not reported in this paper, and experimental data refer to room
temperature. The signal and idler spectra, acquired using a scanning
monochromator with a resolution bandwidth (RBW) of 0.5~nm and a PbSe
photodiode, are reported in Fig.~\ref{spectra}(a) for the poling
periods available from 29.5 to 32.5~$\mu$m. The corresponding signal
and idler wavelengths cover the range 2.25-2.6~$\mu$m and
4.1-4.9~$\mu$m, respectively, with a maximum spectral width of
100~nm at 2.48~$\mu$m. Due to strong water vapor resonances in the
2.55-2.7~$\mu$m range, it was not possible to drive the OPO above
threshold with the remaining poling periods. Figure~\ref{spectra}(b)
shows the average power and pulse duration of the signal and idler
as a function of the central wavelength for each spectrum. The power
levels are in the range 20-60~mW, with a maximum of 60~mW for both
signal and idler when using the 31.5~$\mu$m poling period. The pulse
durations have been measured by an interferometric autocorrelator
based on collinear second harmonic generation in a LiIO$_3$ crystal.
Figure~\ref{spectra}(c) shows the autocorrelation trace of the
signal pulse train with the minimum duration of 110~fs (40-mW
power), as retrieved by a fit assuming a sech$^2$-pulse shape,
corresponding to a time-bandwidth product of 0.46. The reduction of
the signal bandwidth when changing the poling period is due to the
decrease of the phase-matching acceptance bandwidth and to the
reduced transmission of the PPLN crystal (cut-off at
$\sim$5~$\mu$m), which introduces increasing losses on the idler.
This is also confirmed by the trend of the power levels, with the
exception of the data corresponding to the 32.5-$\mu$m poling, where
the reported values are affected by the water vapour absorption at
2.55~$\mu$m. Table~\ref{OPO_performance} summarizes the OPO
performance.

\begin{table}[t]
\caption{OPO performance at room-temperature.}
\label{OPO_performance} \center \footnotesize
\begin{tabular}{cccccccc}
\hline\noalign{\smallskip}

 & threshold & \multicolumn{2}{c}{OPO wavelength} & \multicolumn{2}{c}{output power} & \multicolumn{2}{c}{pulsewidth}\\

poling period & pump power & signal & idler & signal & idler & signal & idler\\

($\mu$m) & (mW) & ($\mu$m) & ($\mu$m) &(mW) & (mW) & (fs) & (fs) \\

\noalign{\smallskip}\hline\noalign{\smallskip}

29.5 & 280 & 2.25 & 4.84 & 20 & 20 & 295 & 400 \\

30.5 & 160 & 2.35 & 4.56 & 24 & 22 & 163 & 510 \\

31.5 & 165 & 2.43 & 4.33 & 60 & 59 & 133 & 492 \\

32.5 & 155 & 2.48 & 4.08 & 37 & 50 & 110 & 470 \\

\noalign{\smallskip}\hline
\end{tabular}
\end{table}

\section{OPO intensity stabilization}

\begin{figure}[t]
\centerline{\includegraphics[width=8.0cm]{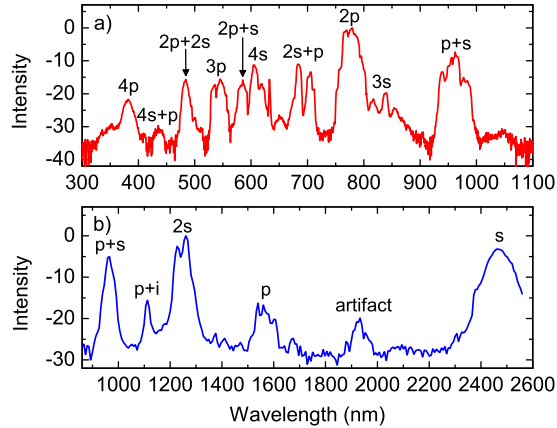}}
\caption{\label{OMA}Nonlinear tones simultaneously generated by the
OPO in the visible (a) and near-infrared (b) spectral region. The
labels have been assigned according to the nonlinear phenomena
corresponding to each tone.}
\end{figure}

\begin{figure}[b]
\vspace{-0.0cm} \centerline{\includegraphics[width=8.0cm]{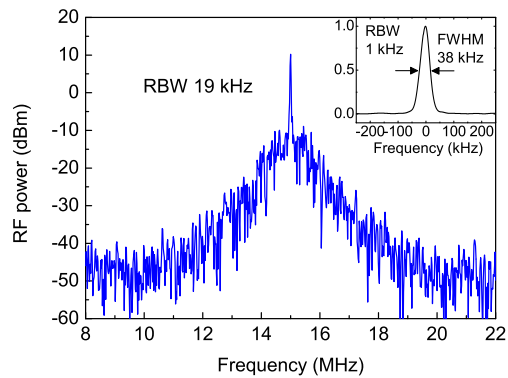}}
\caption{\label{beat}RF spectrum of the beatnote between the $3s$
and $2p$ tones extracted from the OPO cavity, locked to a 15-MHz
reference. Inset: detail of the locked beatnote.}
\end{figure}

The long-term stabilization of the signal intensity has been
achieved by exploiting the wide spectrum of nonlinear mixing signals
generated by non phase-matched interactions inside the PPLN crystal
with a technique similar to that described in~\cite{Adler}.
Figures~\ref{OMA}(a) and (b) show the nonlinear tones extracted from
the HR plane mirror of the OPO cavity in the visible and near
infrared spectral regions, respectively. Each tone has been labelled
according to the specific nonlinear process involved (e.g. $p+i$
represents the sum frequency between the pump and idler comb). The
beam is dispersed using a fused silica prism and focused onto a
125-MHz InGaAs photodiode after proper spectral filtering by a slit.
The spectral overlap around 1200~nm between the components $p+i$ and
$2s$ gives rise to a beatnote at $f_0^p+f_0^i-2f_0^s=2f_0^p-3f_0^s$,
where $f_0^p$, $f_0^s$ and $f_0^i$ are the offset frequency of the
pump, signal and idler, respectively. The same beat frequency has
been observed around 600 and 800~nm, originating from the overlap of
$2p+s$ with $4s$, and $2p$ with $3s$, respectively. A phase-lock
circuit was implemented to stabilize the beatnote against a 15-MHz
reference derived from a frequency synthesizer. The signal at the
output of the InGaAs photodiode has been suitably amplified
($\sim$30~dB) to match the input dynamic of the phase detector, and
then sent to a proportional-integral servo acting on a piezo mounted
cavity mirror. The signal radiation is constituted by frequencies
$\nu_k^s=kv/L=kf_r+f_0^s$, where $k$ is the number of wavelengths
within the cavity, $v$ is the mean phase velocity, $L$ is the cavity
length, and $f_r$ is the repetition frequency. Since the OPO signal
is inherently synchronous with the pump, a change in the cavity
length at constant $f_r$ induces only a change of $f_0^s$
\cite{Gebs}. Therefore, phase-locking of this beat note translates
into an OPO cavity length control, providing long-term stabilization
of the OPO signal intensity.

\begin{figure}[t]
\vspace{-0.0cm} \centerline{\includegraphics[width=10.5cm]{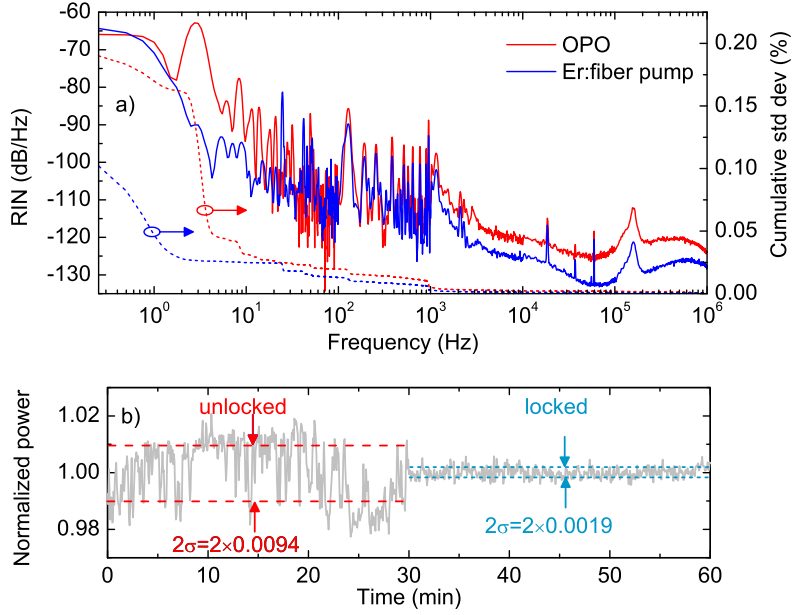}}
\caption{\label{rin} (a) RIN (left) and cumulative standard
deviation (right) of the locked OPO and Er-fiber pump laser. (b)
Signal intensity versus observation time for unlocked (0--30 min)
and locked OPO (30--60 min).}
\end{figure}

Figure~\ref{beat} shows the locked beatnote as measured with an RF
spectrum analyzer. The observed FWHM of 38~kHz is consistent with
the free-running width of the pump offset frequency $f_0^p$, which
is not controlled in this experiment. The beatnote in
Fig.~\ref{beat} has been acquired using the 32.5~$\mu$m poling
period. Similar results in terms of FWHM with slightly reduced peak
values have been obtained for all remaining periods, due to the good
spectral overlap between the $p+i$ and $2s$ nonlinear tones.
Figure~\ref{rin}(a) shows the relative intensity noise (RIN) of the
signal, as recorded by an extended-InGaAs photodiode with 10-MHz
bandwidth, when the OPO cavity length is locked. For comparison, the
RIN of the Er-fiber pump oscillator is also shown. For Fourier
frequencies lower than 20~Hz and larger than 1~kHz an excess noise
in the OPO RIN is observed, due to conversion of the pump offset
frequency phase-noise into OPO amplitude noise
\cite{Ferreiro1,Mulder} and cavity vibrations. The cumulative
standard deviation of the OPO intensity is as low as 0.2\%,
\textit{i.e.} a factor of $\sim$2 larger than the pump laser
cumulative standard deviation of 0.1\%. Long-term intensity
stability of the OPO signal is also shown in Fig.~\ref{rin}(b) for
1~h observation time. The free running standard deviation of the OPO
intensity is effectively reduced by a factor $\sim$5 when the active
stabilization loop of the OPO cavity length is closed.

\section{Conclusion}
\noindent In conclusion, a compact OPO source synchronously pumped
by a commercial 250-MHz Er:fiber laser has been presented. The OPO
signal and idler cover the spectral range 2.25-2.6~$\mu$m and
4.1-4.9~$\mu$m, respectively, with average powers from 20 to 60~mW.
The minimum observed pulse duration is 110~fs at 2.5~$\mu$m, with a
power level of 40~mW. In addition, long-term active stabilization of
the signal intensity in the whole tunability range from 2.25 to
2.6~$\mu$m allows for the reduction of the intensity noise down to
the 0.2\% level. Further stabilization of the pump repetition and
offset frequencies will lead to a full mid-IR comb source, which
could be used for absolute referencing of solid-state Cr$^{2+}$
lasers and quantum cascade lasers, or for direct comb spectroscopy
in the mid-IR spectral region.

\section*{Acknowledgment}
\noindent The authors acknowledge financial support from the Italian
Ministry of University and Research, ELI infrastructure.


\begin{thebibliography}{99}


\bibitem{Diddams} S. A. Diddams,
    ``The evolving optical frequency comb,"
    J. Opt. Soc. Am. B {\bf27,} B51-B62 (2010).

\bibitem{Jiang} J. Jiang, C. Mohr, J. Bethge, M. Fermann, and I. Hartl,
    ``Fully Stabilized, Self-Referenced Thulium Fiber Frequency Comb,"
     in CLEO/Europe and EQEC 2011 Conference Digest, OSA Technical Digest (CD) (OSA, 2011), paper PDB\_1.

\bibitem{Cizmeciyan} M. N. Cizmeciyan, H. Cankaya, A. Kurt, and A. Sennaroglu,
    ``Kerr-lens mode-locked femtosecond Cr$^{2+}$:ZnSe laser at 2420~nm,"
    Opt. Lett. {\bf34,} 3056 (2009).

\bibitem{Neely} T. W. Neely, T. A. Johnson, S. A. Diddams,
``High-power broadband laser source tunable from 3.0~$\mu$m to
4.4~$\mu$, based on a femtosecond Yb:fiber oscillator,"
    Opt. Lett. {\bf36,} 4020 (2011).

\bibitem{Ruehl} A. Ruehl, A. Gambetta, I. Hartl, M. E. Fermann, K. S. E. Eikema, M. Marangoni,
``Widely-tunable mid-IR frequency comb source based on difference
frequency generation," Opt. Lett. \textbf{37,} 2232-2234 (2012).

\bibitem{Adler} F. Adler, K. C. Cossel, M. J. Thorpe, I. Hartl, M. E. Fermann, and J. Ye,
``Phase-stabilized, 1.5~W frequency comb at 2.8–4.8~$\mu$m,"
    Opt. Lett. {\bf34,} 1330 (2009).

\bibitem{Leindecker} N. Leindecker, A. Marandi, R. L. Byer, K. L. Vodopyanov, J. Jiang, I. Hartl, M. Fermann, and P. G. Schunemann,
``Octave-spanning ultrafast OPO with 2.6-6.1~$\mu$m instantaneous
bandwidth pumped by femtosecond Tm-fiber laser,"
    Opt. Express {\bf20,} 7046 (2012).

\bibitem{Washburn} B. R. Washburn, S. A. Diddams, N. R. Newbury, J. W. Nicholson, M. F. Yan, and C. G. Jrgensen,
    ``Phase-locked, erbium-fiber-laser-based frequency comb in the near infrared,"
    Opt. Lett. {\bf29,} 250 (2004).


\bibitem{Leindecker2}N. Leindecker, A. Marandi, R. L. Byer, and K. L. Vodopyanov,
``Broadband degenerate OPO for mid-infrared frequency comb
generation," Opt. Express \textbf{19,} 6296-6302 (2011).

\bibitem{Ferreiro1} T. I. Ferreiro, J. Sun, and D. T. Reid,
``Locking the carrier-envelope-offset frequency of an optical
parametric oscillator without f-2f self-referencing," Opt. Lett.
\textbf{35,} 1668-1670 (2010).

\bibitem{Keilmann} F. Keilmann, C. Gohle, and R. Holzwarth,
``Time-domain mid-infrared frequency-comb spectrometer," Opt. Lett.
\textbf{29,} 1542–1544 (2004).

\bibitem{Bernhardt} B. Bernhardt, A. Ozawa, P. Jacquet, M. Jacquey, Y. Kobayashi, T.
Udem, R. Holzwarth, G. Guelachvili, T. W. H\"ansch, and N. Picque,
``Cavity-enhanced dual-comb spectroscopy," Nat. Photon. \textbf{4,}
55–57 (2009).

\bibitem{Coddington}I. Coddington, W. C. Swann, and N. R. Newbury,
``Coherent multiheterodyne spectroscopy using stabilized optical
frequency combs," Phys. Rev. Lett. \textbf{100,} 013902 (2008).

\bibitem{Haakestad} M. W. Haakestad, H. Fonnum, G. Arisholm, E. Lippert, and K. Stenersen,
    ``Mid-infrared optical parametric oscillator synchronously pumped by an erbium-doped fiber laser,"
    Opt. Express {\bf18,} 25379 (2010).


\bibitem{Gebs} R. Gebs, T. Dekorsy, S. A. Diddams, and A. Bartels
``1-GHz repetition rate femtosecond OPO with stabilized offset
between signal and idler frequency comb,'' Opt. Express \textbf{16,}
5397-5405 (2008).

\bibitem{Mulder} T. D. Mulder, R. P. Scott, and B. H. Kolner,
``Amplitude and envelope phase noise of a modelocked laser predicted
from its noise transfer function and the pump noise power spectrum,"
Opt. Express \textbf{16,} 14186-14191 (2008).


\end{thebibliography}
\end{document}